\documentclass[11pt,leqno]{article}

\usepackage{amsmath}
\usepackage{amsthm}
\usepackage{amssymb}
\usepackage[dvips]{graphicx,color}

\theoremstyle{plain}
\newtheorem{theorem}{Theorem}[section]

\newtheorem{corollary}[theorem]{Corollary}
\theoremstyle{definition}

\theoremstyle{remark}
\newtheorem{remark}[theorem]{Remark}

\numberwithin{equation}{section}

\begin{document}

\title{\textbf{An operator-theoretical study on \\the BCS-Bogoliubov model of superconductivity \\near absolute zero temperature}}
\author{Shuji Watanabe\\
Division of Mathematical Sciences\\
Graduate School of Engineering, Gunma University\\
4-2 Aramaki-machi, Maebashi 371-8510, Japan\\
Email: shuwatanabe@gunma-u.ac.jp}

\date{}

\maketitle

\begin{abstract}
In the preceding papers the present author gave another proof of the existence and uniqueness of the solution to the BCS-Bogoliubov gap equation for superconductivity from the viewpoint of operator theory, and  showed that the solution is partially differentiable with respect to the temperature twice. Thanks to these results, we can indeed partially differentiate the solution and the thermodynamic potential with respect to the temperature twice so as to obtain the entropy and the specific heat at constant volume of a superconductor. In this paper we show the behavior near absolute zero temperature of the thus-obtained entropy, the specific heat, the solution and the critical magnetic field from the viewpoint of operator theory since we did not study it in the preceding papers. Here, the potential in the BCS-Bogoliubov gap equation is an arbitrary, positive continuous function and need not be a constant.

\medskip


\medskip

\end{abstract}

\section{Introduction}

In the BCS-Bogoliubov model of superconductivity, no one gave a proof of the statement that the solution to the BCS-Bogoliubov gap equation is partially differentiable with respect to the temperature. Nevertheless, without such a proof, one partially differentiates the solution and the thermodynamic potential with respect to the temperature twice so as to obtain the entropy and the specific heat at constant volume, and one shows that the phase transition from a normal conducting state to a superconducting state is of the second order. Therefore, if the solution were not partially differentiable with respect to the temperature, then one could not partially differentiate the solution and the thermodynamic potential with respect to the temperature and could not obtain the entropy and the specific heat at constant volume. Moreover, one could not show that the phase transition is of the second order. For this reason, we have to show that the solution is partially differentiable with respect to the temperature twice.

On the basis of fixed-point theorems, the present author \cite[Theorems 2.3 and 2.4]{watanabe-five} (see also \cite[Theorems 2.2 and 2.10]{watanabe-seven}) gave another proof of the existence and uniqueness of the solution and showed that the solution is indeed partially differentiable with respect to the temperature twice. The present author thus showed that the thermodynamic potential is also differentiable with respect to the temperature twice. Here, the potential in the BCS-Bogoliubov gap equation is an arbitrary, positive continuous function and need not be a constant. In this way, the present author gave an operator-theoretical proof of the statement that the phase transition to a superconducting state is of the second order, and solved the long-standing problem of the second-order phase transition from the viewpoint of operator theory. As a result, the present author showed the existence of the first and second order partial derivatives of the solution with respect to the temperature, and showed that all of the solution, the first and second order partial derivatives are continuous functions of both the temperature and the energy. Therefore, thanks to these results, we can indeed differentiate the thermodynamic potential with respect to the temperature twice so as to obtain the entropy and the specific heat at constant volume of a superconductor.

In this paper we show the behavior near absolute zero temperature of the thus-obtained entropy, the specific heat, the solution and the critical magnetic field from the viewpoint of operator theory since we did not study it in the preceding papers \cite{watanabe-five, watanabe-seven}. 

Let $u_0$ be the solution to the BCS-Bogoliubov gap equation \cite{bcs, bogoliubov}, which is a nonlinear integral equation and is given by
\begin{equation}\label{eqn:bcsgapeq}
u_0(T,\,x)=\int_{\varepsilon}^{\hslash\omega_D}
\frac{U(x,\,\xi)\, u_0(T,\, \xi)}{\,\sqrt{\,\xi^2+u_0(T,\, \xi)^2\,}\,}\,
\tanh \frac{\,\sqrt{\,\xi^2+u_0(T,\, \xi)^2\,}\,}{2T}\, d\xi, \   T \geq 0, \   x, \xi \in [\varepsilon,\, \hslash\omega_D].  
\end{equation}
Here, the solution $u_0$ is a function of the absolute temperature $T$ and the energy $x$. The Debye angular frequency $\omega_D$ is a positive constant and depends on a superconductor. The potential $U(\cdot,\,\cdot)$ satisfies $U(x,\,\xi)>0$ at all $(x,\,\xi) \in [\varepsilon, \, \hslash\omega_D]^2$. Throughout this paper we use the unit where the Boltzmann constant $k_B$ is equal to 1.

\begin{remark}
In \eqref{eqn:bcsgapeq} above and \eqref{eqn:thermopotential} below, we introduce a cutoff $\varepsilon>0$ and fix it. If we did not introduce the cutoff $\varepsilon>0$, then the first order derivative of the thermodynamic potential with respect to $T$, and hence the entropy could diverge logarithmically only at  the transition temperature $T_c$. Therefore, the phase transition could not be of the second order. This contradicts a lot of experimental results that the phase transition is of the second order without an external magnetic field. So we introduce the cutoff $\varepsilon>0$. For more details, see \cite[Remarks 1.1, 1.10 and 1.11]{watanabe-seven}.
\end{remark}

\begin{remark}
In the physics literature, one introduces the cutoff $\varepsilon>0$ and avoids the divergence (mentioned in the preceding remark) of the entropy at the transition temperature $T_c$. Then, letting the cutoff tend to 0, one removes the cutoff. From the view point of operator theory, introducing the cutoff $\varepsilon>0$ means that one deals with the Banach space $C([0, \, T_c] \times [\varepsilon,\, \hslash\omega_D])$ (consisting of continuous functions defined on $[0, \, T_c] \times [\varepsilon,\, \hslash\omega_D]$) that the solution $u_0$ to the BCS-Bogoliubov gap equation belongs to. On the other hand, removing the cutoff means that one deals with the Banach space $C([0, \, T_c] \times [0,\, \hslash\omega_D])$. One might think that the former Banach space $C([0, \, T_c] \times [\varepsilon,\, \hslash\omega_D])$ continuously tends to the latter one $C([0, \, T_c] \times [0,\, \hslash\omega_D])$ as the cutoff goes to zero. Note that there is a function that belongs to the former Banach space but not to the latter one. For example, the function $x \mapsto 1/x$ belongs to the former Banach space but not to the latter one. Under this circumstance, unfortunately the present author does not know which norm, which metric (which distance), which $\varepsilon$-neighborhoods, I could use in order to prove the statement that the former Banach space continuously tends to the latter one as the cutoff goes to zero from the view point of operator theory. I therefore introduce the cutoff $\varepsilon>0$, fix it and deal with the former Banach space $C([0, \, T_c] \times [\varepsilon,\, \hslash\omega_D])$.
\end{remark}

For a fixed temperature $T$, the existence and uniqueness of the solution were established and studied in \cite{odeh, billardfano, vansevenant, bls, chen, deugeihailoss, fhns, fhss, freijihaizlseiringer, hhss, hainzlloss, haizlseiringer, haizlseiringer2, haizlseiringer3, watanabe-one, watanabe-two, watanabe-four, watanabe-five, watanabe-seven}. See also Kuzemsky \cite[Chapters 26 and 29]{kuzemsky3} and \cite {kuzemsky, kuzemsky2}. For the role of the chemical potential in the BCS-Bogoliubov model, see Anghel and Nemnes \cite{angnem} and Anghel \cite{ang-one, ang-two}.
 
In connection to this, the BCS-Bogoliubov gap equation plays a role similar to that of the Maskawa--Nakajima equation \cite{maskawa-nakajima-one, maskawa-nakajima-two} which has attracted considerable interest in elementary particle physics. In Professor Maskawa's Nobel lecture, he stated the reason why he dealt with the Maskawa-Nakajima equation. For an operator-theoretical treatment of this equation, see the present author's paper \cite{watanabe-three}.

In the BCS-Bogoliubov model, the thermodynamic potential is given by
\begin{eqnarray}\label{eqn:thermopotential}
\Omega(T)
&=& -2N_0 \int_{\varepsilon}^{\hslash\omega_D} \sqrt{\, x^2+u_0(T,\, x)^2\,} \, dx \\
& & +N_0 \int_{\varepsilon}^{\hslash\omega_D} \frac{u_0(T,\, x)^2}{\,\sqrt{\, x^2+u_0(T,\, x)^2\,}\,}\, \tanh \frac{\,\sqrt{\, x^2+u_0(T,\, x)^2\,}\,}{2T}\, dx \nonumber \\
& & -4N_0 T \int_{\varepsilon}^{\hslash\omega_D} \ln
\left(1+e^{ -\sqrt{\, x^2+u_0(T,\,  x)^2\,}/T } \right) \, dx, \quad 0 \leq T \leq T_c,
 \nonumber
\end{eqnarray}
where $u_0$ is the solution to the BCS-Bogoliubov gap equation \eqref{eqn:bcsgapeq}, $T_c$ is the transition temperature (see \cite[Definition 1.8]{watanabe-seven} for our operator-theoretical definition of $T_c$) and $N_0$ is a positive constant and denotes the density of states per unit energy at the Fermi surface. Here we consider only the contribution from the interval $[-\hslash\omega_D, \, \hslash\omega_D]$, and omit the contribution from the other intervals. In other words, we consider only the contribution from superconductivity. For more details, see \cite[(1.5) and (1.6)]{watanabe-seven}.

As mentioned above, thanks to \cite[Theorems 2.3 and 2.4]{watanabe-five} and \cite[Theorems 2.2 and 2.10]{watanabe-seven}), we can indeed partially differentiate the solution with respect to the temperature $T$ twice, and have the solution $u_0$, the first order partial derivative $\partial u_0/\partial T$ and the second order partial derivative $\partial^2 u_0/\partial T^2$. Moreover, all of them are continuous functions of both the temperature $T$ and the energy $x$. Therefore, thanks to these results, we can indeed differentiate the thermodynamic potential $\Omega$ with respect to $T$ twice so as to obtain the entropy and the specific heat at constant volume. Note that the potential $U(\cdot,\,\cdot)$ in the BCS-Bogoliubov gap equation is an arbitrary, positive continuous function and need not be a constant.

\begin{remark}
If the solution $u_0$ is an accumulating point of the set $V$ in \cite[Theorem 2.2]{watanabe-seven} (resp. of the set $W$ in \cite[Theorem 2.10]{watanabe-seven}), then we replace $u_0$ by a suitably chosen element of $V$ (resp. of $W$) in the form \eqref{eqn:thermopotential} of the thermodynamic potential $\Omega$. This is because  $u_0$ is an accumulating point. Note that such a suitably chosen element is partially differentiable with respect to the temperature $T$ twice and that it is a continuous function of both the temperature $T$ and the energy $x$. Therefore, once we replace the solution $u_0$ by a suitably chosen element in the form \eqref{eqn:thermopotential}, we can differentiate the thermodynamic potential $\Omega$ with respect to the temperature $T$ twice so as to obtain the entropy and the specific heat at constant volume. 
\end{remark}

\section{Main results}

Thanks to Theorem 2.2 in \cite{watanabe-seven}, the solution $u_0$ to the BCS-Bogoliubov gap equation \eqref{eqn:bcsgapeq} satisfies that at all $x \in [\varepsilon,\, \hslash\omega_D]$,
\[
\frac{ \, \partial u_0 \,}{\partial T}(0,\, x) =0 \quad \mbox{and} \quad \frac{\, \partial^2 u_0 \,}{\partial T^2}(0,\, x)=0.
\]
Let $T_0 \, (>0)$ be in a neighborhood of absolute zero temperature $T=0$ and let $(T, \, x) \in [0,\, T_0] \times [\varepsilon,\, \hslash\omega_D]$. Since $\partial u_0/\partial T$ is a continuous function, the value $(\,\partial u_0/\partial T\,)(T,\, x)$ is approximately equal to $(\,\partial u_0/\partial T\,)(0,\, x)$, i.e.,
\[
\frac{ \, \partial u_0 \,}{\partial T}(T,\, x) \approx \frac{ \, \partial u_0 \,}{\partial T}(0,\, x) \, (=0).
\]
The same is true for $\partial^2 u_0/\partial T^2$. Therefore we apply the following approximation.

Approximation (A) \quad  Let $T_0 \, (>0)$ be in a neighborhood of absolute zero temperature $T=0$ and let $(T, \, x) \in [0,\, T_0] \times [\varepsilon,\, \hslash\omega_D]$. Since all of the solution $u_0$, the first order partial derivative $\partial u_0/\partial T$ and the second order partial derivative $\partial^2 u_0/\partial T^2$ are continuous functions of both the temperature $T$ and the energy $x$, we apply the following approximation:
\[
\frac{ \, \partial u_0 \,}{\partial T}(T,\, x) \approx \frac{ \, \partial u_0 \,}{\partial T}(0,\, x) \, (=0), \    \frac{ \, \partial^2 u_0 \,}{\partial T^2}(T,\, x) \approx \frac{\, \partial^2 u_0 \,}{\partial T^2}(0,\, x) \, (=0), \   \frac{(X/T)^n }{\, \cosh (X/T)  \,} \approx 0.
\]
Here, $X>0$ and $n$ is every nonnegative integer.

\begin{remark}
The approximation $u_0(T,\, x) \approx u_0(0,\, x)$ follows from $(\, \partial u_0/ \partial T \,)(0,\, x) =0$ and the approximation $(\, \partial^2 u_0/ \partial T^2 \,)(T,\, x) \approx 0$. Here, $(T, \, x) \in [0,\, T_0] \times [\varepsilon,\, \hslash\omega_D]$.
\end{remark}

\begin{theorem}\label{thm:main} \   Let $u_0$ be the solution to the BCS-Bogoliubov gap equation \eqref{eqn:bcsgapeq} given by Theorem 2.2 in \cite{watanabe-seven}. Suppose Approximation (A) and let $T \in [0,\, T_0]$. Then the thermodynamic potential $\Omega$ is partially differentiable with respect to the temperature $T$ twice, and so there exist the entropy $S$ and the specific heat $C_V$ at constant volume. The entropy $S$, the specific heat $C_V$ at constant volume and the solution $u_0$ are approximated as follows:
\begin{eqnarray*}
S(T) &\approx&
\frac{\, 4N_0 \,}{T} \int_{\varepsilon}^{\hslash\omega_D}
 \sqrt{ \, \xi^2+u_0(0,\, \xi)^2 \, } 
 \exp \left(  -\frac{ \, \sqrt{\,\xi^2+u_0(0,\, \xi)^2 \,} \,}{T} \right) \, d\xi, \\
C_V(T) &\approx& \frac{\, 4N_0 \,}{T^2} \int_{\varepsilon}^{\hslash\omega_D}
 \left\{ \xi^2+u_0(0,\, \xi)^2 \right\}
 \exp \left(  -\frac{ \, \sqrt{\,\xi^2+u_0(0,\, \xi)^2 \,} \,}{T} \right) \, d\xi, \\
u_0(T,\, x) &\approx& u_0(0,\, x)
 -2 \int_{\varepsilon}^{\hslash\omega_D} U(x,\,\xi) \,
   \exp \left(  -\frac{ \, \sqrt{\,\xi^2+u_0(0,\, \xi)^2 \,} \,}{T} \right) \, d\xi.
\end{eqnarray*}
Moreover, the critical magnetic field at absolute zero temperature and the specific heat at the transition temperature $T_c$ satisfy
\[
\frac{H_c(0)^2}{\, T_c \,  C_V(T_c) } =
 \frac{4\pi}{ \displaystyle{ \,  \int_{\varepsilon/(2T_c)}^{\hslash\omega_D/(2T_c)} \frac{\eta^2}{\,  \cosh^2 \eta \,} \, d\eta \, } }
 \int_{\varepsilon/(2T_c)}^{\hslash\omega_D/(2T_c)} 
 \frac{ \{\, \sqrt{\,\eta^2+(2T_c)^{-2} \, u_0(0,\, 2T_c\eta)^2 \,} -\eta\, \}^2 }{\, \sqrt{\,\eta^2+(2T_c)^{-2} \, u_0(0,\, 2T_c\eta)^2 \,} \,} \, d\eta.
\]
\end{theorem}

\begin{remark}
Since $\hslash\omega_D/(2T_c)$ is very large in many superconductors, we often let $\hslash\omega_D/(2T_c) \to \infty$ and $\varepsilon/(2T_c) \to 0$ in the physics literature. 
\end{remark}

\begin{corollary}\label{crl:universal}
Suppose that $u_0(0,\, 2T_c\eta)/T_c$ is a constant and does not depend on superconductors, and let $\hslash\omega_D/(2T_c) \to \infty$ and $\varepsilon/(2T_c) \to 0$. Then $H_c(0)^2/(T_c \,  C_V(T_c))$ does not depend on superconductors and becomes a universal constant.
\end{corollary}

\begin{remark}
As far as the present author knows, similar results are obtained in the physics literature under the restriction that the potential $U(\cdot,\,\cdot)$ in the BCS-Bogoliubov gap equation is a constant. But Theorem \ref{thm:main} holds true even when the potential $U(\cdot,\,\cdot)$ is not a constant but an arbitrary, positive continuous function.
\end{remark}

\begin{remark}
Suppose that the potential $U(\cdot,\,\cdot)$ is a constant, i.e., $U(\cdot,\,\cdot)=U_0$. Here, $U_0$ is a positive constant. Then the solution $u_0$ to the BCS-Bogoliubov gap equation does not depend on the energy $x$ and becomes a function of the temperature $T$ only. We denote the solution by $u_0(T)$. Then the forms of $S(T)$, $C_V(T)$ and $u_0(T, \, x)$ in Theorem \ref{thm:main} are reduced to the following well-known forms, respectively: At $T \in [0,\, T_0]$,
\begin{eqnarray*}
S(T) &\approx& \frac{\, 2\sqrt{2\pi} \, N_0 \, u_0(0)^{3/2} \,}{\sqrt{T}}
 \exp \left(  -\frac{ \, u_0(0) \,}{T} \right), \quad
C_V(T) \approx \frac{\, 2\sqrt{2\pi} \, N_0 \, u_0(0)^{5/2} \,}{T^{3/2}}
 \exp \left(  -\frac{ \, u_0(0) \,}{T} \right), \\
u_0(T) &\approx& u_0(0)-U_0 \sqrt{\, 2\pi \, T\, u_0(0)  \,}
 \exp \left(  -\frac{ \, u_0(0) \,}{T} \right)
\end{eqnarray*}
as $\hslash\omega_D/(2T_c) \to \infty$ and $\varepsilon/(2T_c) \to 0$. The form of $H_c(0)^2/(T_c \,  C_V(T_c))$ in Theorem \ref{thm:main} is reduced to
\[
\frac{H_c(0)^2}{\, T_c \,  C_V(T_c) } =
 \frac{4\pi}{ \displaystyle{ \,  \int_{\varepsilon/(2T_c)}^{\hslash\omega_D/(2T_c)} \frac{\eta^2}{\,  \cosh^2 \eta \,} \, d\eta \, } }
 \int_{\varepsilon/(2T_c)}^{\hslash\omega_D/(2T_c)} \frac{ \{\, \sqrt{\,\eta^2+(2T_c)^{-2} \, u_0(0)^2 \,} -\eta\, \}^2 }{\, \sqrt{\,\eta^2+(2T_c)^{-2} \, u_0(0)^2 \,} \,} \, d\eta.
\]
Therefore, if $u_0(0)/T_c$ does not depend on superconductors, then $H_c(0)^2/(T_c \,  C_V(T_c))$ does not depend on superconductors and becomes a universal constant as $\hslash\omega_D/(2T_c) \to \infty$ and $\varepsilon/(2T_c) \to 0$. Actually, $u_0(0)/T_c$ does not depend on superconductors since
\[
u_0(0)/T_c = 4\, \exp\left[ \, \int_0^{\infty} (\, \ln\eta \,)/(\,\cosh^2\eta  \,) \, d \eta \, \right],
\]
as is shown in the physics literature.
\end{remark}

\section{Proof of Theorem \ref{thm:main}}

We first give a proof for the behavior of the entropy $S$ at $T \in [0,\, T_0]$ in Theorem \ref{thm:main}. Thanks to \cite[Theorems 2.3 and 2.4]{watanabe-five} and \cite[Theorems 2.2 and 2.10]{watanabe-seven}, we can indeed partially differentiate the solution $u_0$ to the BCS-Bogoliubov gap equation with respect to $T$ twice. Therefore, we can also differentiate the thermodynamic potential $\Omega$ with respect to $T$ twice. A straightforward calculation gives

\begin{eqnarray}\label{eqn:one-d-thermopotential}
\frac{\, \partial \Omega \,}{\partial T}(T)
&=& -N_0 \int_{\varepsilon}^{\hslash\omega_D} \frac{1}{\, \sqrt{\,\xi^2+u_0(T,\, \xi)^2\,} \,} \frac{\,\partial u^2 \,}{\partial T}(T, \, \xi) \, d\xi \\
& & +N_0 \int_{\varepsilon}^{\hslash\omega_D}  
 \frac{\,\partial u^2 \,}{\partial T}(T, \, \xi) 
  \frac{1}{\,\sqrt{\,\xi^2+u_0(T,\, \xi)^2\,}\,}
   \tanh \frac{\,\sqrt{\,\xi^2+u_0(T,\, \xi)^2\,}\,}{2T}\, d\xi \nonumber \\
& & -\frac{\, N_0\,}{2} \int_{\varepsilon}^{\hslash\omega_D}  
 \frac{\,\partial u^2 \,}{\partial T}(T, \, \xi) 
  \frac{u_0(T,\, \xi)^2}{\,(\,\xi^2+u_0(T,\, \xi)^2\,)^{3/2}\,}
   \tanh \frac{\,\sqrt{\,\xi^2+u_0(T,\, \xi)^2\,}\,}{2T}\, d\xi \nonumber \\
& & +\frac{\, N_0\,}{\, 4T\,} \int_{\varepsilon}^{\hslash\omega_D}  
 \frac{\,\partial u^2 \,}{\partial T}(T, \, \xi) 
  \frac{u_0(T,\, \xi)^2}{\, \xi^2+u_0(T,\, \xi)^2 \,} 
  \left( \cosh \frac{\,\sqrt{\,\xi^2+u_0(T,\, \xi)^2\,}\,}{2T} \right)^{-2}
   \, d\xi \nonumber \\
& & -\frac{\, N_0\,}{\, 2T^2\,} \int_{\varepsilon}^{\hslash\omega_D}  
  u_0(T,\, \xi)^2 
  \left( \cosh \frac{\,\sqrt{\,\xi^2+u_0(T,\, \xi)^2\,}\,}{2T} \right)^{-2} \nonumber \\
& & -4N_0 \int_{\varepsilon}^{\hslash\omega_D}
 \ln \left(  1+e^{-\sqrt{\,\xi^2+u_0(T,\, \xi)^2\,}/T }\right) \, d\xi \nonumber \\
& & -4N_0 \int_{\varepsilon}^{\hslash\omega_D}
 \frac{1}{\, e^{ \sqrt{\,\xi^2+u_0(T,\, \xi)^2\,}/T } +1  \,}
 \left\{ \frac{\, \sqrt{\,\xi^2+u_0(T,\, \xi)^2\,} \,}{T} \right. \nonumber \\
& & \qquad \qquad \qquad \qquad \qquad \qquad \qquad \qquad \quad
 \left. -\frac{ \frac{\,\partial u^2 \,}{\partial T}(T, \, \xi) }
  { \, 2\sqrt{\,\xi^2+u_0(T,\, \xi)^2\,} \, }  \right\} \, d\xi. \nonumber
\end{eqnarray}

Under Approximation (A), we have
\begin{eqnarray}
S(T) &=& -\frac{\partial \Omega}{\, \partial T \,}(T) \approx
 \frac{\, 4N_0 \,}{T} \int_{\varepsilon}^{\hslash\omega_D}
 \frac{ \sqrt{ \, \xi^2+u_0(0,\, \xi)^2 \, } }{\, e^{\sqrt{\, \xi^2+u_0(0,\, \xi)^2 \,} /T }+1 \, } \, d\xi \\ &\approx& \frac{\, 4N_0 \,}{T} \int_{\varepsilon}^{\hslash\omega_D}
 \sqrt{ \, \xi^2+u_0(0,\, \xi)^2 \, } \, e^{-\sqrt{\, \xi^2+u_0(0,\, \xi)^2 \,} /T } \, d\xi.
 \nonumber
\end{eqnarray}
Note that the sixth term on the right side of \eqref{eqn:one-d-thermopotential} is negligible. This is because the sixth term becomes (at $T \in [0,\, T_0]$)
\[
-4N_0\int_{\varepsilon}^{\hslash\omega_D}
 \ln \left(  1+e^{-\sqrt{\,\xi^2+u_0(T,\, \xi)^2\,}/T }\right) \, d\xi \approx
-4N_0\int_{\varepsilon}^{\hslash\omega_D}
 e^{-\sqrt{\,\xi^2+u_0(0,\, \xi)^2\,}/T } \, d\xi,
\]
which is negligible compared to the seventh term.

We next give a proof for the behavior for the specific heat $C_V$ at constant volume at $T \in [0,\, T_0]$. To this end we differentiate $\partial \Omega/\partial T$ with respect to $T$ again and obtain the second order partial derivative $\partial^2 \Omega/\partial T^2$. The second order partial derivative of the first term on the right side of \eqref{eqn:one-d-thermopotential} becomes
\[
-N_0 \int_{\varepsilon}^{\hslash\omega_D} \left[
\frac{1}{\, \sqrt{\,\xi^2+u_0(T,\, \xi)^2\,} \,} 
 \frac{\,\partial^2 u^2 \,}{\partial T^2}(T, \, \xi)
 -\frac{1}{\, 2 \{ \,\xi^2+u_0(T,\, \xi)^2 \, \}^{3/2} \,} 
  \left\{ \frac{\,\partial u^2 \,}{\partial T}(T, \, \xi) \right\}^2
\right] \, d\xi,
\]
which is approximated by $0$ at $T \in [0,\, T_0]$ under Approximation (A). On the other hand, the second order partial derivative of the last term on the right side of \eqref{eqn:one-d-thermopotential} includes
\[
-\frac{\, 4N_0 \,}{T^3} \int_{\varepsilon}^{\hslash\omega_D}
 \left\{ \, \xi^2+u_0(T,\, \xi)^2 \, \right\}
 \frac{ e^{\sqrt{\, \xi^2+u_0(T,\, \xi)^2 \,}/T} }{\, ( \, e^{ \sqrt{\,\xi^2+u_0(T,\, \xi)^2\,}/T } +1 \, )^2 \,} \, d\xi,
\]
which is the only term that we have at $T \in [0,\, T_0]$ under Approximation (A). We deal with the other terms on the right side of \eqref{eqn:one-d-thermopotential} similarly.

As a result, we obtain under Approximation (A) that (at $T \in [0,\, T_0]$)
\[
\frac{\, \partial^2 \Omega \,}{\partial T^2}(T) \approx -\frac{\, 4N_0 \,}{T^3} \int_{\varepsilon}^{\hslash\omega_D}  \left\{ \, \xi^2+u_0(0,\, \xi)^2 \, \right\}
 e^{-\sqrt{\, \xi^2+u_0(0,\, \xi)^2 \,}/T} \, d\xi.
\]
Therefore, under Approximation (A), we have the following behavior for the specific heat $C_V$ at constant volume at $T \in [0,\, T_0]$:
\begin{eqnarray}
C_V(T) &=& -T \, \frac{\partial^2 \Omega}{\, \partial T^2 \,}(T) \\
&\approx& \frac{\, 4N_0 \,}{T^2} \int_{\varepsilon}^{\hslash\omega_D}
 \left\{ \, \xi^2+u_0(0,\, \xi)^2 \, \right\}
 e^{-\sqrt{\, \xi^2+u_0(0,\, \xi)^2 \,}/T} \, d\xi. \nonumber
\end{eqnarray}

\bigskip

We give a proof for the behavior for the solution $u_0$ at $T \in [0,\, T_0]$. A straightforward calculation gives
\begin{eqnarray*}\label{eq:uzero}
u_0(T,\, x)-u_0(0,\, x) &=&
\int_{\varepsilon}^{\hslash\omega_D} U(x,\,\xi) \, \left\{
 \frac{u_0(T,\, \xi)}{ \, \sqrt{\,\xi^2+u_0(T,\, \xi)^2 \,} \,}
 -\frac{u_0(0,\, \xi)}{ \, \sqrt{\,\xi^2+u_0(0,\, \xi)^2 \,} \,}  \right\} \, d\xi \\
& & -2 \int_{\varepsilon}^{\hslash\omega_D} U(x,\,\xi) \,
   \frac{1}{ \, e^{\frac{ \, \sqrt{\,\xi^2+u_0(T,\, \xi)^2 \,} }{T} }+1 \, } \, d\xi. \nonumber
\end{eqnarray*}
Approximation (A) implies
\[
\frac{u_0(T,\, \xi)}{ \, \sqrt{\,\xi^2+u_0(T,\, \xi)^2 \,} \,}
\]
is approximately equal to
\[
\frac{u_0(0,\, \xi)}{ \, \sqrt{\,\xi^2+u_0(0,\, \xi)^2 \,} \,}.
\]
Therefore, we have
\[
u_0(T,\, x) \approx u_0(0,\, x) -2 \int_{\varepsilon}^{\hslash\omega_D} U(x,\,\xi) \,
   e^{-\frac{ \, \sqrt{\,\xi^2+u_0(0,\, \xi)^2 \,} }{T} } \, d\xi. 
\]

\bigskip

We finally give a proof for the rest of Theorem \ref{thm:main}. Note that Theorem 2.19 (v) in  \cite{watanabe-seven} gives
\[
H_c(0)^2 = 32\pi N_0 T_c^2 \int_{\varepsilon/(2T_c)}^{\hslash\omega_D/(2T_c)}
 \frac{\, \{ \, \sqrt{ \eta^2+(2T_c)^{-2} u_0(0,\, 2T_c\eta)^2 }-\eta \, \}^2  \,}{\, \sqrt{ \eta^2+(2T_c)^{-2} u_0(0,\, 2T_c\eta)^2 } \,} \, d\eta.
\]
Moreover, Lemma 5.2 in  \cite{watanabe-seven} gives
\[
C_V^N(T_c) = 8T_c \int_{\varepsilon/(2T_c)}^{\hslash\omega_D/(2T_c)}
 \frac{N_0 \, \eta^2}{\, \cosh^2 \eta \,} \, d\eta.
\]
Therefore,
\[
\frac{H_c(0)^2}{\, T_c \,  C_V^N(T_c) } =
 \frac{4\pi}{\, \displaystyle{    \int_{\varepsilon/(2T_c)}^{\hslash\omega_D/(2T_c)}   \frac{\eta^2}{\, \cosh^2 \eta \,} \, d\eta  }  \,} \,
 \int_{\varepsilon/(2T_c)}^{\hslash\omega_D/(2T_c)}
 \frac{\, \{ \, \sqrt{ \eta^2+(2T_c)^{-2} u_0(0,\, 2T_c\eta)^2 }-\eta \, \}^2  \,}{\, \sqrt{ \eta^2+(2T_c)^{-2} u_0(0,\, 2T_c\eta)^2 } \,} \, d\eta.
\]
The proof of Theorem \ref{thm:main} is complete.

Corollary \ref{crl:universal} follows immediately from the form $H_c(0)^2/( \, T_c \,  C_V^N(T_c) \,)$ just above.


\noindent \textbf{Contributions}

Shuji Watanabe wrote the main manuscript text and reviewed the manuscript.

\noindent \textbf{Competing interests}

The author declares no competing interests.

\end{document}